\documentclass[12pt]{article}

\usepackage{epsfig,latexsym}
\usepackage{amsmath}

\newcommand{\qed}{\hfill $\Box$}

\newtheorem{defi}{Definition}
\newtheorem{theo}{Theorem}
\newtheorem{lemm}{Lemma}
\newtheorem{coro}{Corollary}

\newtheorem{prop}{Proposition}

\def\inst#1{$^{#1}$}

\title{Maximum Weight Independent Set in $l$Claw-Free Graphs in Polynomial Time}

\author{Andreas Brandst\"adt\inst{1}
\and
Raffaele Mosca\inst{2}}

\begin{document}

\maketitle

\begin{center}
{\footnotesize
 \inst{1} Institut f\"ur Informatik, Universit\"at Rostock, D-18051 Rostock, Germany.\\
\texttt{ab@informatik.uni-rostock.de}

\inst{2} Dipartimento di Economia, Universit\'a degli Studi "G. D'Annunzio", Pescara 65121, Italy.\\
\texttt{r.mosca@unich.it}
}
\end{center}

\begin{center}
{\em ad laudem Domini}
\end{center}

\begin{abstract}
The Maximum Weight Independent Set (MWIS) problem is a well-known NP-hard problem. 
A popular way to study MWIS is to detect graph classes for which MWIS can be solved in polynomial time,
with particular reference to hereditary graph classes, i.e., defined by a hereditary graph property or equivalently by forbidding
one or more induced subgraphs. 

For graphs $G_1, G_2$, $G_1+G_2$ denotes the disjoint union of $G_1$ and $G_2$, and for a constant $l \ge 2$, $lG$ denotes the disjoint union of $l$ copies of $G$.
A {\em claw} has vertices $a,b,c,d$, and edges $ab,ac,ad$. MWIS can be solved for claw-free graphs in polynomial time; the first two polynomial time algorithms were introduced in 1980 by \cite{Minty1980,Sbihi1980}, then revisited by \cite{NakTam2001}, and recently improved by \cite{FaeOriSta2011,FaeOriSta2014}, and by \cite{NobSas2011,NobSas2015} with the best known time bound in \cite{NobSas2015}. Furthermore MWIS can be solved for the following extensions of claw-free graphs in polynomial time: fork-free graphs \cite{LozMil2008}, $K_2$+claw-free graphs \cite{LozMos2005}, and apple-free graphs \cite{BraLozMos2010,BraKleLozMos2008}.

This manuscript shows that for any constant $l$, MWIS can be solved for $l$claw-free graphs in polynomial time. Our approach is based on Farber's approach showing that every $2K_2$-free graph has ${\cal O}(n^2)$ maximal independent sets \cite{Farbe1989}, which directly leads to a polynomial time algorithm for MWIS on $2K_2$-free graphs by dynamic programming. 

Solving MWIS for  $l$claw-free graphs in polynomial time extends known results for claw-free graphs, for $lK_2$-free graphs for any constant $l$ \cite{Aleks1991,FarHujTuz1993,Prisn1995,TsuIdeAriShi1977}, for $K_2$+claw-free graphs, for $2P_3$-free graphs \cite{LozMos2012}, and solves the open questions for $2K_2+P_3$-free graphs and for $P_3$+claw-free graphs being two of the minimal graph classes, defined by forbidding one induced subgraph, for which the complexity of MWIS was an open problem.
\end{abstract}

\section{Introduction}

For any missing notation or reference let us refer to \cite{BraLeSpi1999}.
For a graph $G$, let $V(G)$ ($E(G)$, respectively) denote its vertex set (edge set, respectively). 
For a subset $U \subseteq V(G)$, let $N_G(U) = \{v \in V(G) \setminus U:$ $v$ is adjacent to some
$u \in U\}$ be the {\em neighborhood of U in G}, and $A_G(U) = V \setminus (U \cup N(U))$ be the {\em anti-neighborhood of U in G}.
If $U = \{u_1,\ldots,u_k\}$, then let us simply write $N_G(u_1,\ldots,u_k)$ instead of $N_G(U)$, and $A_G(u_1,\ldots,u_k)$ instead of $A_G(U)$.

For $U \subseteq V(G)$ let $G[U]$ denote the subgraph of $G$ induced by $U$. For a vertex $v \in V(G)$ and for a subset $U \subset V(G)$
(with $v \not \in U$), let us say that {\em $v$ contacts $U$} if $v$ is adjacent to some vertex of $U$, and {\em $v$ dominates $U$} if $v$ is adjacent to each
vertex of $U$. A {\em component of $G$} is the vertex set of a maximal connected subgraph of $G$. 

An {\em independent set} (or a {\em stable set})
of a graph $G$ is a subset of pairwise nonadjacent vertices of $G$. An independent set of $G$ is {\em maximal} if it is not properly contained
in any other independent set of $G$.

For a given graph $H$, a graph $G$ is {\em $H$-free} if none of its induced subgraphs is isomorphic to $H$; in particular, $H$ is called a {\em forbidden induced subgraph of $G$}. Given two graphs $G$ and $F$, $G + F$ denotes the disjoint union of $G$ and $F$; in particular, $2G=G+G$ and in general, for $l \ge 2$, $lG$ denotes the disjoint union of $l$ copies of $G$. 

The following specific graphs are mentioned later. A {\em chordless path} $P_k$ has vertices $v_1,v_2,\ldots,v_k$ and
edges $v_jv_{j+1}$ for $1 \le j < k$. A {\em chordless cycle} $C_k$, $k \ge 4$, has vertices $v_1,v_2,\ldots,v_k$ and edges $v_jv_{j+1}$
for $1 \le j < k$ and $v_kv_1$. A $K_n$ is a complete graph of $n$ vertices. A $K_{1,n}$ is a complete
bipartite graph whose sides respectively have one vertex, called the {\em center} of $K_{1,n}$, and $n$ vertices,
called the {\em leaves} of $K_{1,n}$ (if $n = 1$ then there are two trivial centers). $K_{1,3}$ is also called {\em claw}.

A {\em fork} (sometimes called {\em chair}) has vertices $a,b,c,d,e$, and edges $ab,ac,ad,de$ (thus, a fork contains a claw as an induced subgraph). An {\em apple} is formed by a $C_k$, $k \geq 4$, plus one vertex adjacent to exactly one vertex of the $C_k$. 

For indices $i,j,k \ge 0$, let $S_{i,j,k}$ denote the graph with vertices $u,x_1,\ldots,x_i$, $y_1,\ldots,y_j$, $z_1,\ldots,z_k$ such that the subgraph induced by $u,x_1,\ldots,x_i$ forms a $P_{i+1}$ $(u,x_1,\ldots,x_i)$, the subgraph induced by $u,y_1,\ldots,y_j$ forms a $P_{j+1}$ $(u,y_1,\ldots,y_j)$, and the subgraph induced by $u,z_1,\ldots,z_k$ forms a $P_{k+1}$ $(u,z_1,\ldots,z_k)$, and there are no other edges in $S_{i,j,k}$. Thus, {\em claw} is $S_{1,1,1}$, and $P_k$ is isomorphic to e.g. $S_{0,0,k-1}$.

Let $G$ be a given graph and let $w$ be a weight function on $V(G)$. For an independent set $I$, its weight is $w(I):= \Sigma_{v \in I} w(v)$.   
Let $\alpha_w(G):= \max \{w(I): I$ independent in $G\}$ denote the maximum weight of any independent set of $G$.

The {\em Maximum Weight Independent Set} ({\em MWIS}) problem asks for an independent set  of $G$ of maximum weight.
 
 If all vertices $v$ have the same weight $w(v) = 1$, $\alpha_w(G)=\alpha(G)$ and MWIS is called the {\em MIS} problem.

MWIS is NP-hard \cite{GarJoh1979} and remains NP-hard under various restrictions,
such as for triangle-free graphs \cite{Polja1974} and more generally for graphs without chordless cycle
of given length \cite{Murph1992}, for cubic graphs \cite{GarJoh1977} and more generally for $k$-regular
graphs \cite{FriHedJac1998}, and for planar graphs \cite{GarJohSto1976}. 

It can be solved in polynomial time for various graph classes, such as for $P_4$-free graphs \cite{CorLerSte2004} and more generally
perfect graphs \cite{GroLovSch1984}, for claw-free graphs \cite{FaeOriSta2011,Minty1980,NakTam2001,NobSas2011,Sbihi1980} and
more generally fork-free graphs \cite{Aleks2004/1,LozMil2008} and apple-free graphs \cite{BraLozMos2010,BraKleLozMos2008},
for $2K_2$-free graphs \cite{Farbe1989} and more generally $lK_2$-free graphs for any constant $l$ (by combining an algorithm
generating all maximal independent sets of a graph \cite{TsuIdeAriShi1977} and a polynomial upper bound on the number of
maximal independent sets in $lK_2$-free graphs \cite{Aleks1991,FarHujTuz1993,Prisn1995}), $K_2+$claw-free graphs \cite{LozMos2005},
and $2P_3$-free graphs \cite{LozMos2012}. Furthermore MWIS can be solved in polynomial time for $P_5$-free graphs as recently proved in \cite{LokVatVil2014}.

The first two polynomial time algorithms for MWIS on claw-free graphs were introduced in 1980 by Minty \cite{Minty1980} and independently by Sbihi \cite{Sbihi1980}, then revisited by Nakamura and Tamura \cite{NakTam2001}, and recently improved by Faenza, Oriolo, and Stauffer \cite{FaeOriSta2011,FaeOriSta2014}, and by Nobili and Sassano \cite{NobSas2011,NobSas2015} with the best known time bound in \cite{NobSas2015}.

\begin{theo}\label{theo:claw}{\bf \cite{NobSas2015}}
For claw-free graphs, the MWIS problem can be solved in time $O(n^2 \log n)$. \qed
\end{theo}

Obviously, for every graph $G$ the following holds:

$$\alpha_w(G) = max \{w(v) + \alpha_w(G[A(v)]): v \in V\}$$

Thus, for any graph $G$, MWIS can be reduced to the same problem for the anti-neighborhoods of all vertices of $G$. Then we have:

\begin{prop}\label{K1}
For any graph $F$, if M(W)IS can be solved for $F$-free graphs in polynomial time then M(W)IS can be solved for $K_1 + F$-free graphs in polynomial time. \qed
\end{prop}

Let us report the following result due to Alekseev \cite{Aleks1983,Aleks2004/2}. Let us say that a graph is of {\em type $T$}
if it is graph $S_{i,j,k}$ for some indices $i,j,k$.

\begin{theo}\label{theo:Alekseev}{\bf \cite{Aleks1983}}
Let ${\cal X}$ be a class of graphs defined by a finite set $M$ of forbidden induced subgraphs.
If $M$ does not contain any graph every connected component of which is of type $T$, then the M(W)IS problem is {\em NP}-hard for the class ${\cal X}$.
\end{theo}

Alekseev's result implies that M(W)IS is NP-hard for $K_{1,4}$-free graphs $-$ the fact
that M(W)IS is NP-hard for $K_{1,4}$-free graphs is already mentioned in \cite{Minty1980}.

Unless P = NP, Alekseev's result implies that  for any graph $F$, if M(W)IS is polynomial time solvable for $F$-free graphs,
then each connected component of $F$ is of type $T$. By Proposition \ref{K1}, for any graph $F$,
if M(W)IS can be solved in polynomial time for $F$-free graphs then for any constant $l$, M(W)IS can be solved in polynomial time for $lK_1 + F$-free graphs.
It follows that, since for any constant $l$, M(W)IS can be solved in polynomial time for $lK_2$-free graphs \cite{Aleks1991,FarHujTuz1993,Prisn1995,TsuIdeAriShi1977}, for fork-free graphs \cite{Aleks2004/1,LozMil2008}, for $K_2+$claw-free graphs \cite{LozMos2005}, for $2P_3$-free graphs \cite{LozMos2012}, and for $P_5$-free graphs \cite{LokVatVil2014}, the minimal graphs $F$ of type $T$ for which the complexity of M(W)IS for $F$-free graphs was open are: $P_6$, $S_{1,1,3}$, $S_{1,2,2}$, $K_2+P_4$, $2K_2+P_3$, $P_3$+claw, and thus, the minimal graph classes, defined by forbidding one induced subgraph, for which the complexity of M(W)IS was open are:
\begin{itemize}
\item[ ] $P_6$-free graphs, $S_{1,1,3}$-free graphs, $S_{1,2,2}$-free graphs, $K_2+P_4$-free graphs, $2K_2+P_3$-free graphs, $P_3$+claw-free graphs.
\end{itemize}

In this manuscript, we show that for any constant $l$, MWIS can be solved for $l$claw-free graphs in polynomial time. This extends the known results for MWIS on claw-free graphs,  $lK_2$-free graphs for any constant $l$, $K_2$+claw-free graphs, $2P_3$-free graphs, and solves the open question for MWIS on $2K_2+P_3$-free graphs and on $P_3$+claw-free graphs. 

Our approach is based on Farber's approach showing that every $2K_2$-free graph has ${\cal O}(n^2)$ maximal independent sets \cite{Farbe1989} (reported in Section \ref{2K2fr}), which directly leads to a polynomial time algorithm to solve MWIS for $2K_2$-free graphs by dynamic programming. 

\section{Maximal Independent Sets in $2K_2$-Free Graphs}\label{2K2fr}

In this section let us refer to Algorithm ${\cal A}$ (subsequently called Algorithm Alpha) from \cite{LozMos2005} which formalizes the aforementioned approach by Farber \cite{Farbe1989}; our subsequent approach for MWIS on $l$claw-free graphs is based on this algorithm.

For a $2K_2$-free input graph $G$, Algorithm Alpha produces a family ${\cal S}$ of independent sets of $G$, which can be computed in time 
${\cal O}(n^3)$ and which contains ${\cal O}(n^2)$ members such that each maximal independent set of $G$ is contained in some member of ${\cal S}$.

For a graph $G=(V,E)$ with $|V|=n$, a {\em vertex ordering} $(v_1,v_2,\ldots,v_n)$ of $G$ is a total ordering of the vertex set $V$ of $G$.
For such a vertex ordering $(v_1,v_2,\ldots,v_n)$ of $G$, let $G_i:=G[\{v_1,v_2,\ldots,v_i\}]$ denote the subgraph of $G$ induced by the first $i$ vertices, $i \ge 1$.

Given a vertex ordering $(v_1,v_2,\ldots,v_n)$, at each loop $i$, $1 \le i \le n$, Algorithm Alpha provides a family ${\cal S}_i$ of subsets of $\{v_1,v_2,\ldots,v_i\}$ (by modifying ${\cal S}$ at loop $i$ by extending some of its members or by adding new members) such that each maximal independent set of $G_i$ is contained in some member of ${\cal S}_i$, and finally returns the family ${\cal S}_n = {\cal S}$. 

\noindent {\bf Algorithm Alpha}\\
{\bf Input:} A $2K_2$-free graph $G$ and a vertex ordering $(v_1,v_2,\ldots,v_n)$ of $G$.\\
{\bf Output:} A family ${\cal S}$ of subsets of $V(G)$.\\
\\
${\cal S}:=\{\emptyset\}$;\\
{\bf For} $i=:1$ {\bf to} $n$ {\bf do}\\
{\bf begin}\\
1. [Extension of some members of ${\cal S}$]\\
\hspace*{0.5cm} {\bf For each} $H \in {\cal S}$ {\bf do}\\
\hspace*{1cm} {\bf If} $H \cup \{v_i\}$ is an independent set {\bf then} $H:=H\cup \{v_i\}$.\\
2. [Addition of new members to ${\cal S}$]\\
\hspace*{0.5cm} {\bf For each} $K_2$ of $G_i$ containing $v_i$ (i.e., for each edge $uv_i$ of $G_i$) {\bf do}\\
\hspace*{1cm} $H:=\{v_i\}\cup A_{G_i}(u,v_i)$;\\
\hspace*{1cm} ${\cal S} := {\cal S} \cup \{H\}$.\\
{\bf end.}

Then the MWIS problem can be solved for $2K_2$-free graphs by the following algorithm.

\noindent {\bf Algorithm $2K_2$-Free-MWIS}\\
{\bf Input:} A $2K_2$-free graph $G$.\\ 
{\bf Output:} A maximum weight independent set of $G$.

\begin{itemize}
\item[(1)] Execute Algorithm Alpha for $G$. Let ${\cal S}$ be the resulting family of subsets of $V(G)$.

\item[(2)] For each $H \in {\cal S}$, compute a maximum weight independent set of $G[H]$ (note that each $H \in {\cal S}$ is an independent set since $G$ is $2K_2$-free). Then choose a best solution, i.e., one of maximum weight.
\end{itemize}

Then one obtains the following result.

\begin{theo}{\bf \cite{Farbe1989}}
For $2K_2$-free graphs, the MWIS problem can be solved  in time ${\cal O}(n^{4})$ by Algorithm $2K_2$-Free-MWIS. \qed
\end{theo}

\section{Maximal Independent Sets in Claw+Claw-Free Graphs}\label{2Clawfr}

\subsection{A Basic Lemma}\label{BasicLemma}

First let us introduce a preparatory result.
For each $k \in \{1,\ldots,14\}$, let $L_k$ be the graph drawn in the subsequent figure. 
Note that each $L_k$ contains an induced claw. For each $k \in \{1,\ldots,14\}$, let $W(L_k)$ denote the set of white vertices of $L_k$, let $B(L_k)$ denote
the set of black vertices of $L_k$, and let $top(L_k)$ denote the (white) vertex at the top of $L_k$.

\begin{figure}\label{FigLk}
\centering
\includegraphics[width=\textwidth]{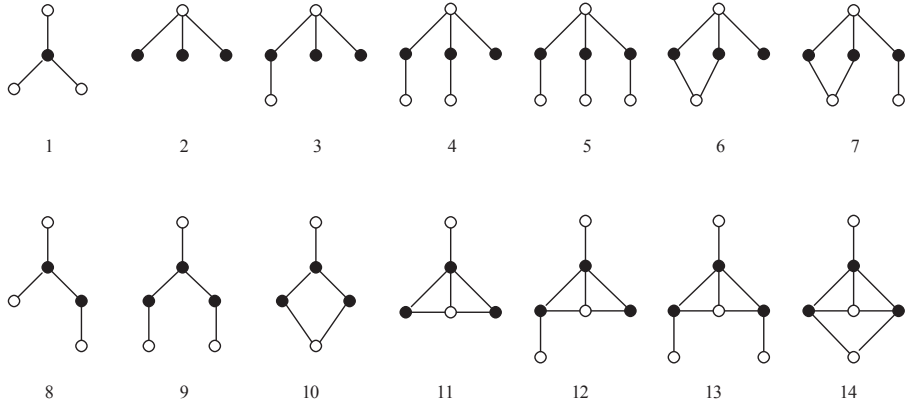}
\caption{Graphs $L_k$ for $k = 1,\ldots,14$}
\end{figure}

\begin{lemm}\label{lemm:claw}
For a graph $G$, assume that $v \in V(G)$ is a vertex such that $v$ is contained in an induced claw of $G$ and $G[V(G) \setminus \{v\}]$ is claw-free.
Then for each maximal independent set $I$ of $G$ with $v \in I$, there is a $k \in \{1,\ldots,14\}$ such that $I \subseteq W(L_k) \cup A_{G}(L_k)$
for an induced subgraph $L_k$ of $G$ with $v$ = top$(L_k)$. 
\end{lemm}

{\bf Proof.} Let $K$ be a claw in $G$ with, say, $V(K)=\{v,a,b,c\}$. Let $I$ be a maximal independent set of $G$ containing $v$, and let $I' := I \setminus \{v\}$. Then for $H:=V(G) \setminus \{v\}$, $I'$ is a maximal independent set of $G[H \setminus N(v)]$. Let us distinguish between the following cases.

{\bf Case 1} $G[H]$ is connected.

By assumption, $v$ is contained in an induced claw of $G$. Let us distinguish between two subcases.

{\bf Case 1.1 $v$ is the center of $K$}.

Since $G[H]$ is claw-free, each of $a,b,c$ has at most two neighbors in $I'$.

{\bf Case 1.1.1} If a vertex of $a,b,c$, say $a$, has two neighbors in $I'$, say $s_1,s_2$ then $I \subseteq W(L_1) \cup A_{G}(L_1)$ with $V(L_1)=\{v,a,s_1,s_2\}$, $W(L_1)=\{v,s_1,s_2\}$, and $v$ = top$(L_1)$.

{\bf Case 1.1.2} If none of $a,b,c$ has a neighbor in $I'$ then $I \subseteq W(L_2) \cup A_{G}(L_2)$ with $V(L_2)=\{v,a,b,c\}$ and $v$ = top$(L_2)$.

{\bf Case 1.1.3} Now assume that Cases 1.1.1 and 1.1.2 are excluded. This means that one of $a,b,c$, say without loss of generality $a$, has exactly one neighbor in $I'$ and $b$ and $c$ have at most one neighbor in $I'$. Let $as_1 \in E$ for $s_1 \in I'$. Note that not both of $b$ and $c$ are adjacent to $s_1$ since $H$ is claw-free, and in general, $a,b$ and $c$ do not have any common neighbor in $I'$. 

If $N(b) \cap I'=N(c) \cap I'=\emptyset$ then we have $I \subseteq W(L_3) \cup A_{G}(L_3)$ with $V(L_3)=\{v,a,b,c,s_1\}$ and $v$ = top$(L_3)$.  

If $b$ has exactly one neighbor in $I'$, say $s_2$, and $N(c) \cap I' = \emptyset$ then 
if $s_1 \neq s_2$, we have $I \subseteq W(L_4) \cup A_{G}(L_4)$ with $V(L_4)=\{v,a,b,c,s_1,s_2\}$ and $v$ = top$(L_4)$, and 
if $s_1 = s_2$, we have $I \subseteq W(L_6) \cup A_{G}(L_6)$ with $V(L_6)=\{v,a,b,c,s_1\}$ and $v$ = top$(L_6)$, and similarly for the case when $c$ has exactly one neighbor in $I'$, and $N(b) \cap I' = \emptyset$.

Finally, assume that both $b$ and $c$ have a neighbor in $I'$, i.e., there are $s_2,s_3 \in I'$ with $bs_2 \in E$ and $cs_3 \in E$.

If $s_1,s_2,s_3$ are pairwise distinct then we have $I \subseteq W(L_5) \cup A_{G}(L_5)$ with $V(L_5)=\{v,a,b,c,s_1,s_2,s_3\}$ and $v$ = top$(L_5)$. 

Now assume that $|\{s_1,s_2,s_3\}|=2$ (recall that $|\{s_1,s_2,s_3\}|=1$ is impossible). Without loss of generality, let $s_1=s_2$. Then we have 
$I \subseteq W(L_7) \cup A_{G}(L_7)$ with $V(L_7)=\{v,a,b,c,s_1,s_3\}$ and $v$ = top$(L_7)$.  

{\bf Case 1.2 $v$ is a leaf of $K$}. 

Without loss of generality, let $b$ be the center of $K$. 
Since $G[H]$ is claw-free, $b$ has at most two neighbors in $I'$, and if $a \notin I'$ ($c \notin I'$, respectively), the same holds for $a$ ($c$, respectively).

The following subcases are exhaustive by symmetry.

{\bf Case 1.2.1} If $a,c \in I'$ then $I \subseteq W(L_1) \cup A_{G}(L_1)$ with $V(L_1)=\{v,a,b,c\}$ and $v$ = top$(L_1)$.

{\bf Case 1.2.2} If exactly one of $a,c$ is in $I'$, say without loss of generality, $a \in I'$ and $c \not \in I'$ (and more generally, only one of the neighbors of $b$ is in $I'$ - otherwise we have Case 1.2.1) then $c$ has a neighbor in $I'$, say $s$, since $I'$ is a maximal independent set of $G[H \setminus N(v)]$. Then clearly, $s$ is nonadjacent to $a$ and $v$ and is nonadjacent to $b$ (otherwise $b$ would have two neighbors in $I'$). Then $I \subseteq W(L_8) \cup A_{G}(L_8)$ 
with $V(L_8)=\{v,a,b,c,s\}$ and $v$ = top$(L_8)$.

{\bf Case 1.2.3} Now assume that Cases 1.2.1 and 1.2.2 are excluded. Thus, $a,c \not \in I'$. Then both $a$ and $c$ must have a neighbor in $I'$ since $I'$ is a maximal independent set of $G[H \setminus N(v)]$. 

If no neighbor of $a$ or $c$ in $I'$ is adjacent to $b$ then both $a$ and $c$ have exactly one neighbor in $I'$,  else a claw in $G[H]$ would arise involving $b$.
Let $s_1,s_2 \in I'$ with $as_1 \in E$, $cs_2 \in E$.

If $s_1 \neq s_2$ then $I \subseteq W(L_9) \cup A_{G}(L_9)$ with $V(L_9)=\{v,a,b,c,s_1,s_2\}$ and $v$ = top$(L_9)$.

If $s_1=s_2$ then $I \subseteq W(L_{10}) \cup A_{G}(L_{10})$ with $V(L_{10})=\{v,a,b,c,s_1\}$ and $v$ = top$(L_{10})$.
 
Now assume that, without loss of generality, a neighbor $s \in I'$ of $a$ is adjacent to $b$. We claim: 

(i) $s$ is adjacent to $c$, since otherwise Case 1.2.2 holds with $s$ instead of $a$; 

(ii) $a$ and $c$ have at most one more neighbor in $I'$, and such a neighbor is non-adjacent to $b$, since otherwise Case 1.2.1 holds (i.e., $b$ has  two neighbors in $I'$).

If neither $a$ nor $c$ have another neighbor in $I'$ then $I \subseteq W(L_{11}) \cup A_{G}(L_{11})$ with $V(L_{11})=\{v,a,b,c,s\}$ and $v$ = top$(L_{11})$.

If there is $s_1 \in I'$ with $s_1 \neq s$, $as_1 \in E$ and the only neighbor of $c$ in $I'$ is $s$ then $I \subseteq W(L_{12}) \cup A_{G}(L_{12})$ with $V(L_{12})=\{v,a,b,c,s,s_1\}$ and $v$ = top$(L_{12})$, and similarly if $c$ has two neighbors $s,s_1 \in I'$ and $a$ has only neighbor $s \in I'$.

Finally, if $a$ and $c$ have another neighbor in $I'$, say $s_1,s_2 \in I'$, $s \neq s_1, s \neq s_2$ with $as_1 \in E$ and $cs_2 \in E$ then we have:

If $s_1 \neq s_2$ then $I \subseteq W(L_{13}) \cup A_{G}(L_{13})$ with $V(L_{13})=\{v,a,b,c,s,s_1,s_2\}$ and $v$ = top$(L_{13})$, and 
if $s_1 = s_2$ then $I \subseteq W(L_{14}) \cup A_{G}(L_{14})$ with $V(L_{14})=\{v,a,b,c,s,s_1\}$ and $v$ = top$(L_{14})$. 

{\bf Case 2}: $G[H]$ is not connected.

This case can be treated similarly as Case 1 in which $G[H]$ is connected. If $G$ is not connected then we can solve MWIS separately for each component of $G$. If $G$ is connected and $v$ is the leaf of a claw then obviously, $G[H]$ is connected. Thus, we can assume that $v$ is the center of a claw $K$, and we can follow the arguments of Case 1. For brevity let us omit the proof, which can be split into the subcases in which $G[H]$ has two or three components.
Finally we have $I \subseteq W(L_k) \cup A_{G}(L_k)$, for some $k \in \{1,\ldots,7\}$, with $v$ = top$(L_k)$.  
\qed

\subsection{MWIS for Claw+Claw-Free Graphs}\label{claw+clawfr}

Now we show that for claw+claw-free graphs, MWIS can be solved in time ${\cal O}(n^{10})$. 
For this, we need the following notion: 
\begin{defi}\label{defi: good family}
Let $G$ be a graph and let ${\cal S}$ be a family of subsets of $V(G)$. Then ${\cal S}$ is a {\em good claw-free family of} $G$ if the following holds: 
\begin{enumerate}
\item[$(i)$] Each member of ${\cal S}$ induces a claw-free subgraph in $G$. 
\item[$(ii)$] Each maximal independent set of $G$ is contained in some member of ${\cal S}$. 
\item[$(iii)$] ${\cal S}$ contains polynomially many members and can be computed in polynomial time.
\end{enumerate}
\end{defi}

The basic step is the subsequent Algorithm Gamma(2) (based on the corresponding Algorithm Alpha of Section \ref{2K2fr}) which, for any claw+claw-free (i.e., 2claw-free) input graph $G$, computes a good claw-free family ${\cal S}$ of $G$. The approach is based on Farber's idea for MWIS on $2K_2$-free graphs described in Algorithm Alpha of Section \ref{2K2fr}.  

{\bf Algorithm Gamma(2)}\\
{\bf Input:} A claw+claw-free graph $G$ and a vertex-ordering $(v_1,v_2,\ldots,v_n)$ of $G$.\\
{\bf Output:} A good claw-free family ${\cal S}$ of $G$.\\

${\cal S}:=\{\emptyset\}$;\\
{\bf For} $i=:1$ {\bf to} $n$ {\bf do}\\
{\bf begin}\\
1. [Extension of some members of ${\cal S}$]\\
\hspace*{0.5cm} {\bf For each} $H \in {\cal S}$ {\bf do}\\
\hspace*{0.5cm} {\bf If} $G[H \cup \{v_i\}]$ is claw-free {\bf then} $H:=H\cup \{v_i\}$.\\ 
2. [Addition of new members to ${\cal S}$]\\
\hspace*{0.5cm} {\bf For each} induced $L_k$ of $G_i$, $k \in \{1,\ldots,14\}$, with $v_i$ = top$(L_k)$ {\bf do}\\
\hspace*{1cm} Compute a good claw-free family, say ${\cal F}$, of $G[A_{G_i}(L_k)]$. \\
\hspace*{1cm} {\bf For each} $F \in {\cal F}$, set ${\cal S}:= {\cal S} \cup \{W(L_k) \cup F\}$. \\ 
{\bf end.}

\begin{prop}\label{prop:claw+claw}
Step $2$ of Algorithm Gamma$(2)$ is well defined, i.e., $G[A_{G_i}(L_k)]$ is claw-free and has a good claw-free family $($formed by one member, namely, 
$A_{G_i}(L_k))$ which can be computed in constant time.
\end{prop}

{\bf Proof.} 
Subgraph $G[A_{G_i}(L_k)]$ is claw-free since $G$ is assumed to be claw+claw-free, each $L_k$ contains an induced claw and $A_{G_i}(L_k)$ is defined as the anti-neighborhood of $V(L_k)$. Then the subgraph $G[A_{G_i}(L_k)]$ has a good claw-free family (formed by one member, namely, $A_{G_i}(L_k)$) which can be computed in constant time.  
\qed

For proving the correctness and the time bound of Algorithm Gamma(2), we need the following lemmas.

\begin{lemm}\label{lemmC:1}
Let $G$ be a claw+claw-free graph and let ${\cal S}$ be the result of Algorithm Gamma$(2)$. Then we have:
\begin{itemize}
	\item[$(i)$] Each member of ${\cal S}$ induces a claw-free subgraph of $G$.
	\item[$(ii)$] Each maximal independent set of $G$ is contained in some member of ${\cal S}$.
\end{itemize}
\end{lemm}

{\bf Proof.} $(i)$: Each member of ${\cal S}$ is created either in the initialization step as the empty set or in Step 1 or Step 2 of some loop. Clearly, each member $H\cup \{v_i\}$ created in Step~1 induces a claw-free subgraph in $G$ since each member of ${\cal S}$ is extended in Step 1 only if the extension preserves its claw-freeness.  
According to Step 2 and to Proposition~\ref{prop:claw+claw}, each member of ${\cal S}$ created in Step 2 is the disjoint union of a vertex subset of a claw-free subgraph, namely $W(L_k)$, and of a claw-free subgraph representing its anti-neighborhood $A_{G_i}(L_k)$, namely a member of a good claw-free family. Therefore, each member of ${\cal S}$ created in Step 2 induces a claw-free graph. This completes the proof of statement $(i)$.

$(ii)$: By ${\cal S}_i$, let us denote the family ${\cal S}$ resulting by the $i$-th loop of Algorithm Gamma(2).
Let us show that for all $i \in \{1,\ldots,n\}$, each maximal independent set of $G_i$ is contained in a member $H$ of ${\cal S}_i$.
The proof is done by induction. For $i=1$, the statement is trivial. Then let us assume that the statement holds for $i-1$ and prove that it holds for $i$.

Let $I$ be a maximal independent set of $G_i$.

If $v_i \not\in I$, then by the induction assumption, $I$ is contained in some member of ${\cal S}_{i-1}$, and thus of ${\cal S}_i$, since each member of ${\cal S}_{i-1}$ is a (not necessarily proper) subset of a member of ${\cal S}_i$.

If $v_i \in I$, then let us consider the following argument. By the induction assumption, let $H \in {\cal S}_{i-1}$ with $I \setminus \{v_i\} \subseteq H$. Note that for all $j$, $1 \le j \le n$, each member of ${\cal S}_j$ induces a claw-free graph, as one can easily verify by an argument similar to the proof of statement $(i)$. Thus, $G[H]$ is claw-free. 

Then let us consider the following two cases which are exhaustive by definition of Algorithm Gamma(2).

{\bf Case 1}: $G[H \cup \{v_i\}]$ is claw-free.

Then $I$ is contained in the set $H \cup \{v_i\}$, which is a member of ${\cal S}_i$ since it is generated by Step 1 of the algorithm at loop $i$.

{\bf Case 2}: $G[H \cup \{v_i\}]$ is not claw-free.

Then by Lemma \ref{lemm:claw}, since $G[H]$ is claw-free, there is a $k \in \{1,\ldots,14\}$ such that $I \subseteq W(L_k) \cup A_{G_i}(L_k)$ for an induced subgraph $L_k$ of $G_i$ with $v_i$ = top$(L_k)$, and $W(L_k) \cup A_{G_i}(L_k)$ is contained in ${\cal S}_i$ since it is generated by Step 2 of Algorithm Gamma(2) at loop $i$. 
\qed

\begin{lemm}\label{lemmC:3}
The family ${\cal S}$ produced by Algorithm Gamma$(2)$ contains ${\cal O}(n^{7})$ members and can be computed in ${\cal O}(n^{9})$ time, which is also the time bound of Algorithm Gamma$(2)$.
\end{lemm}

{\bf Proof.} 
The members of ${\cal S}$ are created either in the initialization step or in Step 2 of all the loops of Algorithm Gamma(2) and then are possibly (iteratively) extended in Step 1 of Algorithm Gamma(2).

Concerning the member created in the initialization step, i.e., the empty set: This member is created in constant time and is possibly (iteratively) extended by Step 1 of each loop in ${\cal O}(n)$ time (and the number of loops is $n$). Then this member can be computed in ${\cal O}(n^{2})$ time.

Concerning the members created in Step 2 of all the loops: Such members are created with respect to all induced $L_k$, $1 \le k \le 14$ (the maximum number of vertices in any $L_k$ is 7), of $G_i$, i.e., with respect to a family of ${\cal O}(n^7)$ subsets of $G_i$ (in fact the algorithm produces the anti-neighborhoods of all $L_k$ for $k \in \{1,\ldots,14\}$ of $G_i$ just once since at loop $i$ all such $L_k$ contain $v_i$). Then for the respective anti-neighborhood, namely $A_{G_i}(L_k)$, of each such subset the algorithm computes a good claw-free family. By Proposition \ref{prop:claw+claw}, $G[A_{G_i}(L_k)]$ is claw-free and has a good claw-free family (which contains one member and can be computed in constant time). Therefore the cardinality of the family of such members is ${\cal O}(n^7)$ and all such members can be created in ${\cal O}(n^7)$ time (since each such member can be created in Step 2 in constant time). Then such members are possibly (iteratively) extended in Step 1 in ${\cal O}(n)$ time (and the number of loops is $n$). Then such members can be computed in ${\cal O}(n^9)$ time.

Therefore, $S$ contains ${\cal O}(n^{7})$ members and can be computed in ${\cal O}(n^{9})$ time, which is also the time bound of Algorithm Gamma(2).  
\qed  

Note that Lemmas \ref{lemmC:1} and \ref{lemmC:3} directly imply the following.

\begin{coro}\label{coro:claw+claw}
Every claw+claw-free graph has a good claw-free family which can be computed by Algorithm Gamma$(2)$.   \qed
\end{coro}

Then the MWIS problem can be solved for claw+claw-free graphs by the following algorithm.

{\bf Algorithm MWIS(2)}\\
{\bf Input:} A claw+claw-free graph $G$.\\ 
{\bf Output:} A maximum weight independent set of $G$.

\begin{itemize}
\item[(1)] Execute Algorithm Gamma(2) for $G$. Let ${\cal S}$ be the resulting family of subsets of $V(G)$.

\item[(2)] For each $H \in {\cal S}$, compute a maximum weight independent set of $G[H]$. Then choose a best solution, i.e., one of maximum weight.
\end{itemize}

\begin{theo}
Algorithm MWIS$(2)$ is correct and can be done in ${\cal O}(n^{10})$ time.
\end{theo}

{\bf Proof.} {\em Correctness}: By Lemma \ref{lemmC:1} $(ii)$, Algorithm MWIS(2) is correct.

{\em Time bound}:
By Lemma \ref{lemmC:3}, step (1) can be executed in ${\cal O}(n^{9})$ time.
By Lemma \ref{lemmC:3}, the family ${\cal S}$ contains ${\cal O}(n^{7})$ members.
Then, by Lemma \ref{lemmC:1} $(i)$ and Theorem \ref{theo:claw}, step (2) can be executed in ${\cal O}(n^{10})$ time.
Thus, Algorithm MWIS(2) can be executed in time ${\cal O}(n^{10})$. 
\qed 

Then one obtains the following result.

\begin{coro}\label{theo:WIS-claw+claw}
For claw+claw-free graphs, the MWIS problem can be solved in time ${\cal O}(n^{10})$ by Algorithm MWIS$(2)$. \qed
\end{coro}

\section{MWIS for $l$Claw-Free Graphs}\label{MWISlclaw-fr}

In this section we show that for any fixed $l \ge 2$, MWIS for $l$claw-free graphs can be solved in polynomial time.
For this, we first describe the subsequent Algorithm Gamma($l$), which for any $l$claw-free input graph $G$ computes a good claw-free family ${\cal S}$ of $G$. The approach recursively uses Algorithm Gamma$(l-1)$ for Algorithm Gamma($l$), starting with Algorithm Gamma(2) of subsection \ref{claw+clawfr}.  

{\bf Algorithm Gamma$(l)$}\\
{\bf Input:} An $l$claw-free graph $G$ and a vertex-ordering $(v_1,v_2,\ldots,v_n)$ of $G$.\\
{\bf Output:} A good claw-free family $S$ of $G$.\\

${\cal S}:=\{\emptyset\}$;\\
{\bf For} $i=:1$ {\bf to} $n$ {\bf do}\\
{\bf begin}\\
1. [Extension of some members of ${\cal S}$]\\
\hspace*{0.5cm} {\bf For each} $H \in {\cal S}$ {\bf do}\\
\hspace*{0.5cm} {\bf If} $G[H \cup \{v_i\}]$ is claw-free {\bf then} $H:=H\cup \{v_i\}$.\\ 
2. [Addition of new members to ${\cal S}$]\\
\hspace*{0.5cm} {\bf For each} induced $L_k$ of $G_i$, $k \in \{1,\ldots,14\}$, with $v_i$ = top$(L_k)$ {\bf do}\\
\hspace*{1cm} Compute a good claw-free family, say ${\cal F}$, of $G[A_{G_i}(L_k)]$ by\\ 
\hspace*{1cm} Algorithm Gamma$(l-1)$. \\
\hspace*{1cm} {\bf For each} $F \in {\cal F}$, set ${\cal S}:= {\cal S} \cup \{W(L_k) \cup F\}$. \\ 
{\bf end.}

{\bf Assumption 1.} To prove the subsequent Proposition \ref{prop:lclaw}, Lemmas \ref{lemmC:1l} and \ref{lemmC:3l}, and Corollary \ref{coro:lclaw}, we need to consider them as a {\em unique} result, in order to give a proof by induction on $l$. For $l$ = 2, the proof of Proposition \ref{prop:lclaw}, of Lemmas \ref{lemmC:1l} and \ref{lemmC:3l}, and of Corollary \ref{coro:lclaw} is respectively that of Proposition \ref{prop:claw+claw}, of Lemmas \ref{lemmC:1} and \ref{lemmC:3}, and of Corollary \ref{coro:claw+claw}. 

Then let us assume that the subsequent Proposition \ref{prop:lclaw}, Lemmas \ref{lemmC:1l} and \ref{lemmC:3l}, and Corollary \ref{coro:lclaw} hold for $l-1$ and let us show that they hold for $l$.

\begin{prop}\label{prop:lclaw}
Step $2$ of Algorithm Gamma$(l)$ is well defined, i.e., $G[A_{G_i}(L_k)]$ is $(l-1)$claw-free and has a good claw-free family which can be computed by Algorithm 
Gamma$(l-1)$.
\end{prop}

{\bf Proof.} 
Subgraph $G[A_{G_i}(L_k)]$ is $(l-1)$claw-free since $G$ is $l$claw-free and since $A_{G_i}(L_k)$ is defined as the anti-neighborhood of $L_k$ containing an induced claw. Then by Assumption 1, i.e., by Corollary \ref{coro:lclaw} with respect to $l-1$, subgraph $G[A_{G_i}(L_k)]$ has a good claw-free family which can be computed by Algorithm Gamma$(l-1)$.  
\qed

For proving the correctness and the time bound of Algorithm Gamma$(l)$, we need the following lemmas.

\begin{lemm}\label{lemmC:1l}
Let $G$ be an $l$claw-free graph and let ${\cal S}$ be the result of Algorithm Gamma$(l)$. Then we have:
\begin{itemize}
	\item[$(i)$] Each member of ${\cal S}$ induces a claw-free subgraph of $G$.
	\item[$(ii)$] Each maximal independent set of $G$ is contained in some member of ${\cal S}$.
\end{itemize}
\end{lemm}

{\bf Proof.} 
According to Assumption~1, the proof is similar to that of Lemma~\ref{lemmC:1}, with Proposition~\ref{prop:lclaw} instead of Proposition~\ref{prop:claw+claw} and with Algorithm Gamma($l$) instead of Algorithm Gamma(2).  
\qed

\begin{lemm}\label{lemmC:3l}
The family ${\cal S}$ produced by Algorithm Gamma$(l)$ contains polynomially many members and can be computed in polynomial time, which is also the time bound of Algorithm Gamma$(l)$.
\end{lemm}

{\bf Proof.} 
The members of ${\cal S}$ are created either in the initialization step or in Step 2 of all the loops of Algorithm Gamma($l$) and then are possibly (iteratively) extended in Step 1 of Algorithm Gamma($l$).

Concerning the member created in the initialization step, i.e., the empty set, this member is created in constant time and is possibly (iteratively) extended by Step 1 of each loop in ${\cal O}(n)$ time (the number of loops is $n$). Then this member can be computed in ${\cal O}(n^{2})$ time.

Concerning the members created in Steps 2 of all the loops, such members are created with respect to all induced $L_k$ of $G$, i.e., with respect to a family of ${\cal O}(n^7)$ subsets of $G$ (in fact, the algorithm produces the anti-neighborhoods of all $L_k$ for $k \in \{1,\ldots,14\}$ of $G_i$ just once since at loop $i$ all such $L_k$ contain $v_i$ as their top vertex). Then for the respective anti-neighborhood, namely $A_{G_i}(L_k)$, of each such subset the algorithm computes a good claw-free family. By Proposition \ref{prop:lclaw}, $G[A_{G_i}(L_k)]$ is $(l-1)$claw-free and has a good claw-free family. Therefore the cardinality of the family of such members is bounded by a polynomial and all such members can be created in polynomial time (since each such member can be created in Step~2 in polynomial time). Then such members are possibly (iteratively) extended in Step~1 in ${\cal O}(n)$ time (the number of loops is $n$). Thus, such members can be computed in polynomial time.

Therefore, ${\cal S}$ can be computed in polynomial time, which is also the time bound of Algorithm Gamma($l$).  \qed

Note that Lemmas \ref{lemmC:1l} and \ref{lemmC:3l} directly imply the following.

\begin{coro}\label{coro:lclaw}
For any fixed $l$, each $l$claw-free graph has a good claw-free family which can be computed via Algorithm Gamma$(l)$.  \qed
\end{coro}

Then for $l$claw-free graphs, the MWIS problem can be solved by the following algorithm. \\

{\bf Algorithm MWIS($l$)}\\
{\bf Input:} An $l$claw-free graph $G$.\\ 
{\bf Output:} A maximum weight independent set of $G$.

\begin{itemize}
\item[(1)] Execute Algorithm Gamma$(l)$ for $G$. Let ${\cal S}$ be the resulting family of subsets of $V(G)$.

\item[(2)] For each $H \in {\cal S}$, compute a maximum weight independent set of $G[H]$. Then choose a best solution, i.e., one of maximum weight.
\end{itemize}

\begin{theo}
Algorithm MWIS$(l)$ is correct and can be executed in polynomial time.
\end{theo}

{\bf Proof.} {\em Correctness}: By Lemma \ref{lemmC:1l} $(ii)$, Algorithm MWIS$(l)$ is correct.

{\em Time bound}: By Lemma \ref{lemmC:3l}, step (1) can be executed in polynomial time.
By Lemma~\ref{lemmC:3l}, the family ${\cal S}$ contains polynomially many members.
Then by Lemma \ref{lemmC:1l} $(i)$ and by Theorem \ref{theo:claw}, step (2) can be executed in polynomial time.
Thus, Algorithm MWIS$(l)$ can be executed in polynomial time. 
\qed 

Then one obtains the following result.

\begin{coro}\label{theo: WIS-lclaw}
For any fixed $l$, the MWIS problem can be solved in polynomial time for $l$claw-free graphs by Algorithm MWIS$(l)$. \qed
\end{coro}

\begin{footnotesize}
\renewcommand{\baselinestretch}{0.4}

\end{footnotesize}

\end{document}